\documentclass[12pt]{article}%
\usepackage{amsmath,latexsym}
\usepackage{graphicx}
\usepackage{amsmath}
\usepackage{amsfonts}
\usepackage{amssymb}%
\begin{document}

\title{{Dark-energy wormholes in generalized 
    Kaluza-Klein gravity}}
   \author{
Peter K. F. Kuhfittig*\\  \footnote{kuhfitti@msoe.edu}
 \small Department of Mathematics, Milwaukee School of
Engineering,\\
\small Milwaukee, Wisconsin 53202-3109, USA}

\date{}
 \maketitle

\begin{abstract}\noindent
This paper discusses traversable
wormholes in a dark-energy setting by starting with
a model due to Sung-Won Kim.  This model, based on
the Friedmann-Lema\^{i}tre-Robertson-Walker model, is
combined with a generalized Kaluza-Klein model.
It is well known that phantom dark energy can in
principle support traversable wormholes due to the
violation of the null energy condition (NEC), a 
necessary condition for holding a wormhole open.
Since the phantom divide constitutes a serious
barrier, we retain the dark-energy assumption
$a''(t)>0$ but stay below the phantom divide.  To
show that the violation of the NEC can only come
from the fifth dimension requires a careful analysis
of the scale factor $a(t)$.  While needed in the
overall model, certain 1-forms lack this scale factor
since the extra dimension, often assumed to be
compactified, would not be affected by the
cosmological expansion.  The result is a viable
wormhole model of the Sung-Won-Kim type.  Our main
conclusion involves Morris-Thorne wormholes: due
to the Kaluza-Klein gravity, we obtain a valid
wormhole solution without crossing the phantom divide.
\end{abstract}

\section{Introduction}\label{S:Introduction}
Wormholes are handles or tunnels connecting widely separated
regions of our Universe or entirely different universes.
That such structures might be suitable for interstellar
travel was first proposed by Morris and Thorne \cite{MT88}
by means of the following static and spherically symmetric
line element for the wormhole spacetime:
\begin{equation}\label{E:L1}
ds^{2}=-e^{2\Phi(r)}dt^{2}+\frac{dr^2}{1-b(r)/r}
+r^{2}(d\theta^{2}+\text{sin}^{2}\theta\,
d\phi^{2}),
\end{equation}
using units in which $c=G=1$.   Here $b=b(r)$ is called
the \emph{shape function} and $\Phi =\Phi(r)$ is called
the \emph{redshift function}, which must be everywhere
finite to prevent the occurrence of an event horizon. For
the shape function, we must have $b(r_0)=r_0$, where
$r=r_0$ is the radius of the \emph{throat} of the
wormhole.  An important requirement is the \emph{flare-out
condition} at the throat: $b'(r_0)\le 1$, while $b(r)
<r$ near the throat.  The flare-out condition can only
be met by violating the null energy condition (NEC),
which states that
\begin{equation}
   T_{\alpha\beta}k^{\alpha}k^{\beta}\ge 0\,\,
   \text{for all null vectors}\,\, k^{\alpha},
\end{equation}
where $T_{\alpha\beta}$ is the energy-momentum tensor.
In particular, for the outgoing null vector $(1,1,0,0)$,
the violation takes on the form
\begin{equation}\label{E:null}
   T_{\alpha\beta}k^{\alpha}k^{\beta}=\rho+p_r<0.
\end{equation}
Here $T^t_{\phantom{tt}t}=-\rho$ is the
energy density, $T^r_{\phantom{rr}r}= p_r$
is the radial pressure, and
$T^\theta_{\phantom{\theta\theta}\theta}=
T^\phi_{\phantom{\phi\phi}\phi}=p_t$ is
the lateral (transverse) pressure.

In a cosmological setting, we may consider the
equation of state of a perfect fluid, $p=\omega\rho$,
where $p=p_r=p_t$.  Referring to the
Friedmann-Lema\^{i}tre-Robertson-Walker (FLRW) model,
\begin{equation}\label{E:L2}
  ds^2=-dt^2+[a(t)]^2\left(\frac{dr^2}{1-kr^2}+r^2
  (d\theta^2+\text{sin}^2\theta\,d\phi^2)\right),
\end{equation}
the Friedmann equation
\begin{equation}
   \frac{\ddot{a}(t)}{a(t)}=-\frac{4\pi}{3}
   (\rho+3p)=-\frac{4\pi}{3}(\rho+3\omega\rho)
\end{equation}
implies that we must have $\omega<-\frac{1}{3}$
to yield an accelerated expansion, commonly
referred to as ``dark energy."  The special case
$\omega=-1$ corresponds to Einstein's cosmological
constant, while $\omega<-1$ is usually referred
to as phantom dark energy.  Here $\rho+p=\rho+
\omega\rho=\rho(1+\omega)<0$, so that the NEC
is automatically violated.  (In other words, we 
already know that phantom dark energy can in 
principle support traversable wormholes.)  In 
the equation of state $p=\omega\rho$, going 
from $\omega>-1$ to $\omega<-1$ is often 
referred to as ``crossing the phantom divide" 
since we are up against a natural barrier.  As 
a result, we will assume an accelerated expansion 
due to dark energy but avoid crossing the phantom 
barrier.  So the source of the NEC violation would 
have to originate elsewhere, in our case, the extra
spatial dimension from the Kaluza-Klein model.
This is our main result.

\emph{Remark:} Regarding the violation of the 
NEC, other proposals have been made.  For example, 
it was shown by Lobo and Oliveira \cite{LO09} that
$f(R)$ modified gravity allows a wormhole to be
constructed from ordinary matter, while the
unavoidable violation of the NEC can be
attributed to the higher-order curvature terms,
interpreted as a gravitational fluid.  Another
possibility, proposed in Refs. \cite{RKRI12}
and \cite{pK15}, is an appeal to noncommutative
geometry, an offshoot of string theory, to account
for the violation.

\section{The cosmological models}
 Our actual starting point is going to be a
 cosmological model due to S.-W. Kim \cite{Kim96},
 based on the FLRW model
\begin{equation}\label{E:L3}
  ds^2=-e^{2\Phi(r)}dt^2+[a(t)]^2\left(\frac{dr^2}
  {1-kr^2-\frac{b(r)}{r}}+r^2
  (d\theta^2+\text{sin}^2\theta\,d\phi^2)\right),
\end{equation}
but combined with the Kaluza-Klein model:
\begin{equation}\label{E:L4}
  ds^2=-e^{2\Phi(r)}dt^2+[a(t)]^2\left(\frac{dr^2}
  {1-kr^2-\frac{b(r)}{r}}+r^2
  (d\theta^2+\text{sin}^2\theta\,d\phi^2)
  +e^{2\Psi(r)}dq^2\right).
\end{equation}
Traversable wormholes sustained by an extra spatial
dimension are also discussed in Ref. \cite{pK10},
which provides a motivation for the form of the
extra term: the model is intended to be as general
as possible, even allowing the fifth dimension to
be compactified.  The embedding of four-dimensional
spacetimes in higher-dimensional flat spacetimes is
discussed in Refs. \cite{sM19, sM16, sM17}.  Paul
Wesson \cite{pW15} has shown that the field equations
in a five-dimensional flat space yield the Einstein
field equations in four dimensions, also called
\emph{the induced-matter theory}.

In line element (\ref{E:L4}), $a(t)$ is the usual
scale factor, where $k$ is the sign of the curvature
of spacetime, i.e., $k=+1, 0, \text{or} -1$.  The
shape function $b=b(r)$ is assumed to have the usual
properties, but we need to keep in mind that meeting
the flare-out condition does not imply a violation
of the NEC, unlike the case of a Morris-Thorne
wormhole; the violation must therefore be confirmed
separately.

Since the scale factor $a(t)$ is one of the issues
discussed in this paper, let us first consider the
following alternative to Eq. (\ref{E:L4}):
\begin{equation}\label{E:L5}
  ds^2=-e^{2\Phi(r)}dt^2+[a(t)]^2\left(\frac{dr^2}
  {1-kr^2-\frac{b(r)}{r}}+r^2
  (d\theta^2+\text{sin}^2\theta\,d\phi^2)
  \right)+e^{2\Psi(r)}dq^2.
\end{equation}
The ever-growing $a(t)$  causes the last term to
become negligible.  Our interest in wormholes,
however, draws our attention to the local behavior
operating on much smaller scales.  While not
absolutely necessary, we make the usual
assumption that the extra dimension is small or
even compactified.  So locally, such a small
dimension would not be affected by the
cosmological expansion, clearly indicating
that Eq. (\ref{E:L4}) is the appropriate model
on a cosmological scale.  We will discussed this
further in the next section [after Eq. (\ref{E:omit})].
Our main goal in that section is to obtain the
tools for yielding complete wormhole solutions.
Because of the sheer generality of the solutions,
we will not be able to discuss the physical
implications until we reach Sections \ref{S:initial}
and \ref{S:final}.

\section{Primary calculations}

To obtain a wormhole solution based on the
cosmological model (\ref{E:L4}), we need to
choose an orthonormal basis $\{e_{\hat{\alpha}}\}$
that is dual to the following 1-form basis:
\begin{equation}
    \theta^0=e^{\Phi(r)}\, dt,
\end{equation}
\begin{equation}
    \theta^1=a(t)\left[1-kr^2-\frac{b(r)}{r}
    \right]^{-1/2}\,dr,
\end{equation}
\begin{equation}
   \theta^2=a(t)\,r\,d\theta,
\end{equation}
\begin{equation}
      \theta^3=a(t)\,r\,
\,\text{sin}\,\theta\,d\phi,
\end{equation}
\begin{equation}
    \theta^4=a(t)\,e^{\Psi(r)}dq.
\end{equation}
These forms yield
\begin{equation}
   dt=e^{-\Phi(r)}\,\theta^0,
\end{equation}
\begin{equation}\label{E:dr}
   dr=\frac{1}{a(t)}\left[1-kr^2-
   \frac{b(r)}{r}\right]^{1/2}\,\theta^1,
\end{equation}
\begin{equation}
   d\theta=\frac{1}{a(t)}\frac{1}{r}\theta^2,
\end{equation}
\begin{equation}
    d\phi=\frac{1}{a(t)}
    \frac{1}{r\,\text{sin}\,\theta}\theta^3,
\end{equation}
\begin{equation}\label{E:omit}
   dq=e^{-\Psi(r)}\,\theta^4.
\end{equation}
The last equation does not contain the scale factor $a(t)$
since the extra curled-up dimension would not be affected
by the expansion of the Universe, already discussed in the
previous section.  (Further comments regarding $a(t)$ can
be found in Appendix A.)

Our main goal in this section is to obtain the components
of the Riemann curvature tensor.  Here we use the method of
differential forms, following Ref. \cite{HT90}.  To obtain
the curvature 2-forms, we need to calculate the following
exterior derivatives in terms of $\theta^i$, where $b=b(r)$:
\begin{equation}
   d\theta^0=\frac{1}{a(t)}\frac{d\Phi(r)}{dr}
   \left(1-kr^2-\frac{b(r)}{r}\right)
   ^{1/2}\,\theta^1\wedge\theta^0,
\end{equation}
\begin{equation}
   d\theta^1=\frac{a'(t)}{a(t)}e^{-\Phi(r)}
   \,\theta^0\wedge\theta^1,
\end{equation}
\begin{equation}
    d\theta^2=\frac{1}{a(t)}\frac{1}{r}\left(1-kr^2-\frac{b(r)}{r}\right)^{1/2}
    \theta^1\wedge\theta^2
    +\frac{a'(t)}{a(t)}e^{-\Phi(r)} \theta^0\wedge\theta^2,
\end{equation}
\begin{equation}
     d\theta^3=\frac{1}{a(t)}\frac{1}{r}\left(1-kr^2-\frac{b(r)}{r}\right)^{1/2}
     \theta^1\wedge\theta^3+\frac{a'(t)}{a(t)}e^{-\Phi(r)}\theta^0\wedge\theta^3
     +\frac{1}{a(t)}\frac{1}{r}\text{cot}\,\theta
     \,\,\theta^2\wedge\theta^3,
\end{equation}
\begin{equation}
   d\theta^4=\frac{d\Psi(r)}{dr}
   \left(1-kr^2-\frac{b(r)}{r}\right)^{1/2}
   \,\theta^1\wedge\theta^4+a'(t)\,e^{-\Phi(r)}\,\theta^0\wedge\theta^4.
\end{equation}
The connection 1-forms $\omega^i_{\phantom{i}\,\,k}$ have the
symmetry
    $\omega^0_{\phantom{i}\,\,i}=\omega^i_{\phantom{0}0}
    \;(i=1,2,3,4)$\;\text{and}\;$\omega^i_{\phantom{j}j}=
     -\omega^j_{\phantom{i}\,i}\;(i,j=1,2,3,4, i\ne j)$,
and are related to the basis $\theta^i$ by
\begin{equation}
   d\theta^i=-\omega^i_{\phantom{k}k}\wedge\theta^k.
\end{equation}
The solution of this system is found to be
\begin{equation}
   \omega^0_{\phantom{0}1}=\frac{1}{a(t)}\frac{d\Phi(r)}{dr}
   \left(1-kr^2-\frac{b(r)}{r}\right)^{1/2}\theta^0
      +\frac{a'(t)}{a(t)}e^{-\Phi(r)}\theta^1,
\end{equation}
\begin{equation}
   \omega^2_{\phantom{0}1}=\frac{1}{a(t)}
   \frac{1}{r}\left(1-kr^2-\frac{b(r)}{r}\right)^{1/2}\theta^2,
\end{equation}
\begin{equation}
    \omega^3_{\phantom{0}1}=\frac{1}{a(t)}
    \frac{1}{r}\left(1-kr^2-\frac{b(r)}{r}\right)^{1/2}\theta^3,
\end{equation}
\begin{equation}
   \omega^3_{\phantom{0}2}=\frac{1}{a(t)}
   \frac{1}{r}\,\text{cot}\,\theta\,\,\theta^3,
\end{equation}
\begin{equation}\label{E:omega}
   \omega^4_{\phantom{0}1}=
   \frac{d\Psi(r)}{dr}
      \left(1-kr^2-\frac{b(r)}{r}\right)^{1/2}\theta^4,
\end{equation}
\begin{equation}
   \omega^2_{\phantom{0}0}=\frac{a'(t)}{a(t)}e^{-\Phi(r)}\theta^2,
\end{equation}
\begin{equation}
   \omega^3_{\phantom{0}0}=\frac{a'(t)}{a(t)}e^{-\Phi(r)}\theta^3,
\end{equation}
\begin{equation}\label{E:40}
   \omega^4_{\phantom{0}0}=a'(t)\,e^{-\Phi(r)}\theta^4,
\end{equation}
\begin{equation}\label{E:zero}
   \omega^2_{\phantom{0}4}=
   \omega^3_{\phantom{0}4}=0.
\end{equation}

The curvature 2-forms $\Omega^i_{\phantom{j}j}$ are calculated
directly from the Cartan structural equations
\begin{equation}
    \Omega^i_{\phantom{j}j}=d\omega^i_{\phantom{j}j} +\omega^i
     _{\phantom{j}k}\wedge\omega^k_{\phantom{j}j}.
\end{equation}
These forms are listed in Appendix A.

The components of the Riemann curvature tensor can be read off
directly from the form
\begin{equation}
   \Omega^i_{\phantom{j}j}=-\frac{1}{2}R_{mnj}^{\phantom{mnj}i}
    \;\theta^m\wedge\theta^n
\end{equation}
and are listed next.
\begin{multline}
   R_{011}^{\phantom{000}0}=\frac{1}{[a(t)]^2}\left[
   \frac{d^2\Phi(r)}{dr^2}\left(1-kr^2-\frac{b(r)}{r}\right)
   +\left(\frac{d\Phi(r)}{dr}\right)^2\left(1-kr^2-\frac{b(r)}{r}\right)\right.\\
   \left.-\frac{1}{2}\frac{d\Phi(r)}{dr}\left(2kr+\frac{rb'(r)-b(r)}{r^2}\right)\right]
   -\frac{a''(t)}{a(t)}e^{-2\Phi(r)},
\end{multline}
\begin{equation}
   R_{022}^{\phantom{000}0}=R_{033}^{\phantom{000}0}=\frac{1}{[a(t)]^2}
   \frac{1}{r}\frac{d\Phi(r)}{dr}\left(1-kr^2-\frac{b(r)}{r}\right)
   -\frac{a''(t)}{a(t)}e^{-2\Phi(r)},
\end{equation}
\begin{equation}
R_{122}^{\phantom{000}0}=R_{133}^{\phantom{000}0}=
   \frac{a'(t)}{[a(t)]^2}e^{-\Phi(r)}\frac{d\Phi(r)}{dr}
   \left(1-kr^2-\frac{b(r)}{r}\right)^{1/2},
\end{equation}
\begin{equation}
   R_{044}^{\phantom{000}0}=\frac{1}{a(t)}\frac{d\Phi(r)}{dr}
     \frac{d\Psi(r)}{dr}\left(1-kr^2-\frac{b(r)}{r}\right)
     -a''(t)\,e^{-2\Phi(r)},
\end{equation}
\begin{equation}
   R_{144}^{\phantom{000}0}=\frac{a'(t)}{a(t)}e^{-\Phi(r)}\frac{d\Phi(r)}{dr}
   \left(1-kr^2-\frac{b(r)}{r}\right)^{1/2},
\end{equation}
\begin{equation}
   R_{122}^{\phantom{000}1}=R_{133}^{\phantom{000}1}=\frac{1}{[a(t)]^2}
   \frac{1}{2r}\left(-2kr-\frac{rb'(r)-b(r)}{r^2}\right)
   -\left(\frac{a'(t)}{a(t)}\right)^2e^{-2\Phi(r)},
\end{equation}
\begin{multline}
   R_{144}^{\phantom{000}1}=\frac{1}{a(t)}\left[\left(\frac{d\Psi(r)}{dr}\right)^2
   \left(1-kr^2-\frac{b(r)}{r}\right)+\frac{d^2\Psi(r)}{dr^2}\left(1-kr^2-\frac{b(r)}{r}\right)
   \right.\\ \left.+\frac{d\Psi(r)}{dr}\cdot\frac{1}{2}\left(-2kr-\frac{rb'(r)-b(r)}{r^2}\right)\right]
   -\frac{[a'(t)]^2}{a(t)}e^{-2\Phi(r)},
\end{multline}
\begin{equation}
   R_{233}^{\phantom{000}2}=-\frac{1}{[a(t)]^2}\frac{1}{r^2}
   \left(kr^2+\frac{b(r)}{r}\right)-
   \left(\frac{a'(t)}{a(t)}\right)^2e^{-2\Phi(r)},
\end{equation}
\begin{equation}
   R_{244}^{\phantom{000}2}=R_{344}^{\phantom{000}3}=
   \frac{1}{a(t)}\frac{1}{r}\frac{d\Psi(r)}{dr}\left(1-kr^2-\frac{b(r)}{r}\right)
   -\frac{[a'(t)]^2}{a(t)}e^{-2\Phi(r)}.
\end{equation}

The last form to be derived in this section is the Ricci
tensor, which is obtained by a trace on the Riemann
curvature tensor:
\begin{equation}
   R_{ab}=R_{acb}^{\phantom{000}c}.
\end{equation}
The components are listed in Appendix B.

\section{Sung-Won-Kim wormholes}
     \label{S:initial}

Returning now to line elements (\ref{E:L3}) and
(\ref{E:L4}), one way to continue is to follow a
procedure introduced by S.-W. Kim \cite{Kim96}:
separate the Einstein field equations into two
parts, the cosmological part and the wormhole part.
This procedure has been used in Refs. \cite{CAB13}
and \cite{pK17}.  Since we are primarily interested
in Morris-Thorne wormholes, we will eventually let
$k=0$, which turns out to be sufficient for our
main result.  Since the forms in the previous 
section are all in the orthonormal frame, we can use 
a simpler notation: $T_{00}=\rho$ is the energy 
density and $T_{11}=p_r$ is the radial pressure, 
as noted after Eq. (\ref{E:null}).  So
\begin{equation}\label{E:exotic}
   8\pi (\rho +p_r)=[R_{00}-\frac{1}{2}R(-1)]+[R_{11}-
   \frac{1}{2}R(1)]\\=R_{00}+R_{11}.
\end{equation}
We now get directly from Appendix B that
\begin{multline}
  8\pi(\rho +p_r)=-2\frac{1}{[a(t)]^2}\cdot\frac{1}{2r}
         \left(-2kr-\frac{rb'(r)-b(r)}{r^2}\right)\\
         -\frac{1}{a(t)}\left\{\left[\Psi'(r)\right]^2
         \left(1-kr^2-\frac{b(r)}{r}\right)
         +\Psi''(r)\left(1-kr^2-\frac{b(r)}{r}\right) 
         \right.\\\left.
         +\Psi'(r)\cdot\frac{1}{2}\left(-2kr-\frac{rb'(r)-b(r)}{r^2}
         \right)\right\}\\
         +e^{-2\Phi(r)}\left[-2\frac{a''(t)}{a(t)}-a''(t)
         +2\left(\frac{a'(t)}{a(t)}\right)^2+\frac{[a'(t)]^2}{a(t)}
          \right]\\
          +\frac{1}{a(t)}\Phi'(r)\Psi'(r)\left(1-kr^2-\frac{b(r)}{r}\right)
          +2\cdot\frac{1}{[a(t)]^2}\frac{1}{r}\Phi'(r)
          \left(1-kr^2-\frac{b(r)}{r}\right).
\end{multline}
Given the size of the scale factor $a(t)$, the first and last
terms on the right-hand side can be considered negligible,
while in the third term, $-a''(t)$ is dominant and negative
(since $a''(t)>0$).  It follows that $8\pi(\rho +p_r)<0$ for
the proper choices of $\Phi(r)$ and $\Psi(r)$, provided that
$k\le 0$, corresponding to an open Universe.  A simple example
is $\Psi(r)\equiv \text{constant}$.  So the NEC has
been violated without crossing the phantom barrier.  The result
is a wormhole of the Sung-Won-Kim type.

\section{Morris-Thorne wormholes in a dark-energy setting}
    \label{S:final}
Since we are primarily interested in a dark-energy setting,
let us now assume that $\Phi(r)\equiv 0$, as in the FLRW model.
To obtain a Morris-Thorne wormhole, we let $k=0$.  Omitting
the negligible terms, we now get
\begin{multline}\label{E:NEC}
   8\pi(\rho +p_r)=-a''(t)-\frac{1}{a(t)}\left[[\Psi'(r)]^2
   \left(1-\frac{b(r)}{r}\right)
   +\Psi''(r)\left(1-\frac{b(r)}{r}\right)
   \right.\\\left.
    +\Psi'(r)\cdot\frac{1}{2}\left(
   -\frac{rb'(r)-b(r)}{r^2}\right)
   \right].
\end{multline}
Recall that while $a''(t)>0$, we would like to refrain
from crossing the phantom barrier.  Choosing $\Psi(r)$
properly (there are many choices besides 
$\Psi(r)\equiv \text{constant}$) shows that our 
Kaluza-Klein model is sufficiently general to yield 
$\rho +p_r<0$.  So a dark-energy background can in 
principle support Morris-Thorne wormholes thanks to 
the extra spatial dimension.

As another illustration, suppose we consider a similar 
model,
\begin{equation}
  ds^2=-e^{2\Phi(r)}dt^2+[a(t)]^2\left(\frac{dr^2}
  {1-kr^2-\frac{b(r)}{r}}+r^2
  (d\theta^2+\text{sin}^2\theta\,d\phi^2)
  +(1-kr^2)dq^2\right),
\end{equation}
based on the discussions in Refs. \cite{uM16} and
\cite{RR19}.  In view of Eq. (\ref{E:L4}), $\Psi(r)
=\text{ln}\,(1-kr^2)$.  So in the vicinity of the 
throat $r=r_0$, for $k$ sufficiently small, 
Eq. (\ref{E:NEC}) implies that 
$8\pi(\rho +p_r)|_{r=r_0}\approx -a''(t)<0$.


\section{Conclusion}
This paper starts with a wormhole model due to S.-W. Kim,
based on the FLRW model, and is then combined with a
generalized Kaluza-Klein model.  It is well known that
phantom dark energy can in principle support a traversable
wormhole due to the automatic violation of the NEC.  Since
the phantom divide constitutes a serious barrier, we
assume that $a''(t)>0$ while remaining below the phantom
barrier.  The violation can only come from the extra
spatial dimension, but this calls for a careful analysis
of the scale factor $a(t)$: while required in the overall
model, certain 1-forms lack the factor $a(t)$ since the
extra dimension, often assumed to be compactified, is not
affected by the cosmological expansion.  The result is a
viable wormhole model of the Sung-Won-Kim type provided
that $k\le 0$. For a Morris-Thorne wormhole, we assume
that $k=0$ and that $\Phi(r)\equiv 0$ for the redshift
function; the latter ensures consistency with the FLRW
model.  The result is a traversable wormhole that avoids
the phantom barrier.  The violation of the NEC can be
attributed to the extra spatial dimension.

As a final comment, the wormholes considered here can
only exist on very large scales.  See, for example, Ref.
\cite{pK22} and references therein.
\\
\\
\\
\textbf{APPENDIX A\quad The curvature 2-forms}\\
From the Cartan structural equations
\begin{equation*}
    \Omega^i_{\phantom{j}j}=d\omega^i_{\phantom{j}j} +\omega^i
     _{\phantom{j}k}\wedge\omega^k_{\phantom{j}j},
\end{equation*}
we obtain
\begin{multline*}
\Omega^0_{\phantom{0}1}=\frac{1}{[a(t)]^2}\left[\frac{1}{2}\frac{d\Phi(r)}{dr}
  \left(2kr+\frac{rb'(r)-b(r)}{r^2}\right)-\frac{d^2\Phi(r)}{dr^2}
  \left(1-kr^2-\frac{b(r)}{r}\right)\right.\\
     \left.-\left(\frac{d\Phi(r)}{dr}\right)^2\left(1
     -kr^2-\frac{b(r)}{r}\right)\right]\theta^0\wedge\theta^1
     +\frac{a''(t)}{a(t)}e^{-2\Phi(r)}\theta^0\wedge\theta^1
\end{multline*}
\begin{multline*}
   \Omega^0_{\phantom{0}2}=-\frac{1}{[a(t)]^2}\frac{1}{r}\frac{d\Phi(r)}{dr}
   \left(1-kr^2-\frac{b(r)}{r}\right)\theta^0\wedge\theta^2
   +\frac{a''(t)}{a(t)}e^{-2\Phi(r)}\theta^0\wedge\theta^2\\
   -\frac{a'(t)}{[a(t)]^2}e^{-\Phi(r)}\frac{d\Phi(r)}{dr}
   \left(1-kr^2-\frac{b(r)}{r}\right)^{1/2}\theta^1\wedge\theta^2
\end{multline*}
\begin{multline*}
   \Omega^0_{\phantom{0}3}=-\frac{1}{[a(t)]^2}\frac{1}{r}\frac{d\Phi(r)}{dr}
   \left(1-kr^2-\frac{b(r)}{r}\right)\theta^0\wedge\theta^3
   +\frac{a''(t)}{a(t)}e^{-2\Phi(r)}\theta^0\wedge\theta^3\\
   -\frac{a'(t)}{[a(t)]^2}e^{-\Phi(r)}\frac{d\Phi(r)}{dr}
   \left(1-kr^2-\frac{b(r)}{r}\right)^{1/2}\theta^1\wedge\theta^3
\end{multline*}
\begin{multline*}
  \Omega^1_{\phantom{0}2}=-\frac{1}{[a(t)]^2}\cdot\frac{1}{2r}
   \left(-2kr-\frac{rb'(r)-b(r)}{r^2}\right) \theta^1\wedge\theta^2
   +\left(\frac{a'(t)}{a(t)}\right)^2e^{-2\Phi(r)}\,\theta^1\wedge\theta^2\\
   +\frac{a'(t)}{[a(t)]^2}e^{-\Phi(r)}\frac{d\Phi(r)}{dr}\left(1-kr^2-\frac{b(r)}{r}\right)^{1/2}
   \theta^0\wedge\theta^2
\end{multline*}
\begin{multline*}
    \Omega^1_{\phantom{0}3}=-\frac{1}{[a(t)]^2}\cdot\frac{1}{2r}
   \left(-2kr-\frac{rb'(r)-b(r)}{r^2}\right) \theta^1\wedge\theta^3
   +\left(\frac{a'(t)}{a(t)}\right)^2e^{-2\Phi(r)}\,\theta^1\wedge\theta^3\\
   +\frac{a'(t)}{[a(t)]^2}e^{-\Phi(r)}\frac{d\Phi(r)}{dr}\left(1-kr^2-\frac{b(r)}{r}\right)^{1/2}
   \theta^0\wedge\theta^3
\end{multline*}
\begin{equation*}
   \Omega^2_{\phantom{0}3}=\frac{1}{[a(t)]^2}\frac{1}{r^2}
   \left(kr^2+\frac{b(r)}{r}\right)\,\theta^2\wedge\theta^3
   +\left(\frac{a'(t)}{a(t)}\right)^2e^{-2\Phi(r)}\,\theta^2\wedge\theta^3
\end{equation*}
\begin{equation*}
   \Omega^2_{\phantom{0}4}=-\frac{1}{a(t)}\frac{1}{r}\frac{d\Psi(r)}{dr}
   \left(1-kr^2-\frac{b(r)}{r}\right)\theta^2\wedge\theta^4
   +\frac{[a'(t)]^2}{a(t)}e^{-2\Phi(r)}\theta^2\wedge\theta^4
\end{equation*}
\begin{equation*}
    \Omega^3_{\phantom{0}4}=-\frac{1}{a(t)}\frac{1}{r}\frac{d\Psi(r)}{dr}
   \left(1-kr^2-\frac{b(r)}{r}\right)\theta^3\wedge\theta^4
   +\frac{[a'(t)]^2}{a(t)}e^{-2\Phi(r)}\theta^3\wedge\theta^4
\end{equation*}

The Cartan structural equations
    $\Omega^i_{\phantom{j}j}=d\omega^i_{\phantom{j}j} +\omega^i
     _{\phantom{j}k}\wedge\omega^k_{\phantom{j}j}$ show
that $\Omega^0_{\phantom{0}4}$ and $\Omega^1_{\phantom{0}4}$
are the only 2-forms that involve the differentials
$d\omega^0_{\phantom{0}4}$ and $d\omega^1_{\phantom{0}4}$;
this is significant since $\omega^0_{\phantom{0}4}$ and
$\omega^1_{\phantom{0}4}$ are the only nonzero 1-forms
pertaining to the extra spatial dimension. [See Eqs.
(\ref{E:omega}), (\ref{E:40}), and (\ref{E:zero}).]  As noted
in the Introduction, the fifth dimension is not affected by
the cosmological expansion.  So $\theta^4=e^{\Psi(r)}dq$ for
this case.  To clarify these comments, let us briefly consider
the calculation
\begin{multline*}
  d\omega^0_{\phantom{0}4}=d\left[a'(t)\,e^{-\Phi(r)}\theta^4\right]
  =d\left[a'(t)\,e^{-\Phi(r)}e^{\Psi(r)}dq\right]\\
  =a'(t)\left[e^{-\phi(r)}e^{\Psi(r)}\Psi'(r)dq\right]
  =a'(t)\left[e^{-\Phi(r)}e^{\Psi(r)}\Psi'(r)-e^{-\Phi(r)}\Phi'(r)
  \right]dr\wedge dq.
\end{multline*}
After substituting Eqs. (\ref{E:dr}) and (\ref{E:omit}) and adding
the only nonzero term $\omega^0_{\phantom{0}1}\wedge \omega^1_{\phantom{0}4}$,
we obtain the final form
\begin{multline*}
   \Omega^0_{\phantom{0}4}=-\frac{1}{a(t)}\Phi'(r)\Psi'(r)
   \left(1-kr^2-\frac{b(r)}{r}\right)\theta^0\wedge\theta^4
   +a''(t)\,e^{-2\Phi(r)}\theta^0\wedge\theta^4\\
   -\frac{a'(t)}{a(t)}\left(1-kr^2-\frac{b(r)}{r}\right)^{1/2}
   e^{-\Phi(r)}\Phi'(r)\theta^1\wedge\theta^4.
\end{multline*}

The 2-form $\Omega^1_{\phantom{0}4}$ is obtained similarly:
\begin{multline*}
   \Omega^1_{\phantom{0}4}=-\frac{1}{a(t)}\left[
   \left[\Psi'(r)\right]^2\left(1-kr^2-\frac{b(r)}{r}\right)
   +\Psi''(r)\left(1-kr^2-\frac{b(r)}{r}\right)
   +\frac{1}{2}\Psi'(r)\left(-2kr-\frac{rb'(r)-b(r)}{r^2}\right)
   \right]\theta^1\wedge\theta^4\\
   +\frac{a'(t)}{a(t)}e^{-\Phi(r)}\Phi'(r)\left(1-kr^2-\frac{b(r)}{r}\right)^{1/2}
   \theta^0\wedge\theta^4
   +\frac{[a'(t)]^2}{a(t)}e^{-2\Phi(r)} \theta^1\wedge\theta^4.
\end{multline*}
\\
\\
\textbf{APPENDIX B\quad The components of the Ricci tensor}
\begin{multline*}
  R_{00}=\frac{1}{[a(t)]^2}\left[\Phi''(r)\left(1-kr^2-\frac{b(r)}{r}\right)
  +\left[\Phi'(r)\right]^2\left(1-kr^2-\frac{b(r)}{r}\right)
  -\frac{1}{2}\Phi'(r)\left(2kr+\frac{rb'(r)-b(r)}{r^2}\right)\right]\\
  -\frac{a''(t)}{a(t)}e^{-2\Phi(r)}+2\frac{1}{[a(t)]^2}\frac{1}{r}\Phi'(r)
  \left(1-kr^2-\frac{b(r)}{r}\right)-2\frac{a''(t)}{a(t)}e^{-2\Phi(r)}\\
  -a''(t)\,e^{-2\Phi(r)}+\frac{1}{a(t)}\Phi'(r)\Psi'(r)
  \left(1-kr^2-\frac{b(r)}{r}\right)
\end{multline*}
\begin{multline*}
   R_{11}=-\frac{1}{[a(t)]^2}\left[\Phi''(r)\left(1-kr^2-\frac{b(r)}{r}\right)
  +\left[\Phi'(r)\right]^2\left(1-kr^2-\frac{b(r)}{r}\right)
  -\frac{1}{2}\Phi'(r)\left(2kr+\frac{rb'(r)-b(r)}{r^2}\right)\right]\\
  +\frac{a''(t)}{a(t)}e^{-2\Phi(r)}-2\frac{1}{[a(t)]^2}\frac{1}{2r}
  \left(-2kr-\frac{rb'(r)-b(r)}{r^2}\right)
  +2\left(\frac{a'(t)}{a(t)}\right)^2e^{-2\Phi(r)}\\
  -\frac{1}{a(t)}\left[\left[\Psi'(r)\right]^2\left(1-kr^2-\frac{b(r)}{r}\right)
  +\Psi''(r)\left(1-kr^2-\frac{b(r)}{r}\right)+\Psi'(r)\cdot\frac{1}{2}
  \left(-2kr-\frac{rb'(r)-b(r)}{r^2}\right)\right]\\
     +\frac{[a'(t)]^2}{a(t)}e^{-2\Phi(r)}
\end{multline*}
\begin{multline*}
   R_{22}=R_{33}=-\frac{1}{[a(t)]^2}\frac{1}{r}\Phi'(r)
   \left(1-kr^2-\frac{b(r)}{r}\right)+\frac{a''(t)}{a(t)}e^{-2\Phi(r)}\\
   -\frac{1}{[a(t)]^2}\cdot\frac{1}{2}\frac{1}{r}
   \left(-2kr-\frac{rb'(r)-b(r)}{r^2}\right)\\
   +\left(\frac{a'(t)}{a(t)}\right)^2e^{-2\Phi(r)}+\frac{1}{[a(t)]^2}
   \frac{1}{r^2}\left(kr^2+\frac{b(r)}{r}\right)
   +\left(\frac{a'(t)}{a(t)}\right)^2e^{-2\Phi(r)}\\
   -\frac{1}{a(t)}\frac{1}{r}\Psi'(r)\left(1-kr^2-\frac{b(r)}{r}\right)
   +\frac{[a'(t)]^2}{a(t)}e^{-2\Phi(r)}
\end{multline*}
\begin{multline*}
  R_{44}=a''(t)\,e^{-2\Phi(r)}-\frac{1}{a(t)}\Phi'(r)\Psi'(r)
  \left(1-kr^2-\frac{b(r)}{r}\right)\\-\frac{1}{a(t)}\left[
  \left[\Psi'(r)\right]^2\left(1-kr^2-\frac{b(r)}{r}\right)
  +\Psi''(r)\left(1-kr^2-\frac{b(r)}{r}\right)\right.\\\left.
  +\Psi'(r)\cdot\frac{1}{2}\left(-2kr-\frac{rb'(r)-b(r)}{r^2}
  \right)\right]\\
  +\frac{[a'(t)]^2}{a(t)}e^{-2\Phi(r)}+2\left[-\frac{1}{a(t)}
  \frac{1}{r}\Psi'(r)\left(1-kr^2-\frac{b(r)}{r}\right)
  +\frac{[a'(t)]^2}{a(t)}e^{-2\Phi(r)}\right]
\end{multline*}

\end{document}